\documentclass[preprint,journal]{vgtc}       

\ifpdf
  \pdfoutput=1\relax                   
  \usepackage{graphicx}                
  \DeclareGraphicsExtensions{.pdf,.png,.jpg,.jpeg} 
\else
  \usepackage{graphicx}                
  \DeclareGraphicsExtensions{.eps}     
\fi%

\graphicspath{{figures/}{pictures/}{images/}{./}} 

\usepackage{microtype}                 
\PassOptionsToPackage{warn}{textcomp}  
\usepackage{textcomp}                  
\usepackage{mathptmx}                  
\usepackage{times}                     
\usepackage{cite}                      
\usepackage{tabu}                      
\usepackage{booktabs}                  
\usepackage[inline]{enumitem}
\usepackage{fixmath}  
\usepackage[autostyle]{csquotes}
\usepackage{xcolor}
\usepackage[colorinlistoftodos]{todonotes}

\ieeedoi{10.1109/TVCG.2019.2934266}

\onlineid{1171}

\vgtccategory{Research}
\vgtcpapertype{system}

\title{VASSL: A Visual Analytics Toolkit for Social Spambot Labeling}

\author{Mosab Khayat, Morteza Karimzadeh, Jieqiong Zhao, David S. Ebert, \textit{Fellow, IEEE}}
\authorfooter{
\item Mosab Khayat, Jieqiong Zhao and David S. Ebert are with Purdue University. Email: \{mkhayat, zhao413, ebertd\}@purdue.edu.
\item Morteza Karimzadeh is with the University of Colorado Boulder (formerly at Purdue University). Email: karimzadeh@colorado.edu.
}

\shortauthortitle{Khayat \MakeLowercase{\textit{et al.}}:Visual Analytics for Social Spambot Detection}

\abstract{Social media platforms are filled with social spambots. Detecting these malicious accounts is essential, yet challenging, as they continually evolve to evade detection techniques. In this article, we present VASSL, a visual analytics system that assists in the process of detecting and labeling spambots. Our tool enhances the performance and scalability of manual labeling by providing multiple connected views and utilizing dimensionality reduction, sentiment analysis and topic modeling, enabling insights for the identification of spambots. The system allows users to select and analyze groups of accounts in an interactive manner, which enables the detection of spambots that may not be identified when examined individually. We present a user study to objectively evaluate the performance of VASSL users, as well as capturing subjective opinions about the usefulness and the ease of use of the tool.
} 

\keywords{Spambot, Labeling, Detection, Visual Analytics, Social Media Annotation}

\CCScatlist{ 
 \CCScat{K.6.1}{Management of Computing and Information Systems}%
{Project and People Management}{Life Cycle};
 \CCScat{K.7.m}{The Computing Profession}{Miscellaneous}{Ethics}
}

\teaser{
  \centering
  \includegraphics[width=\linewidth]{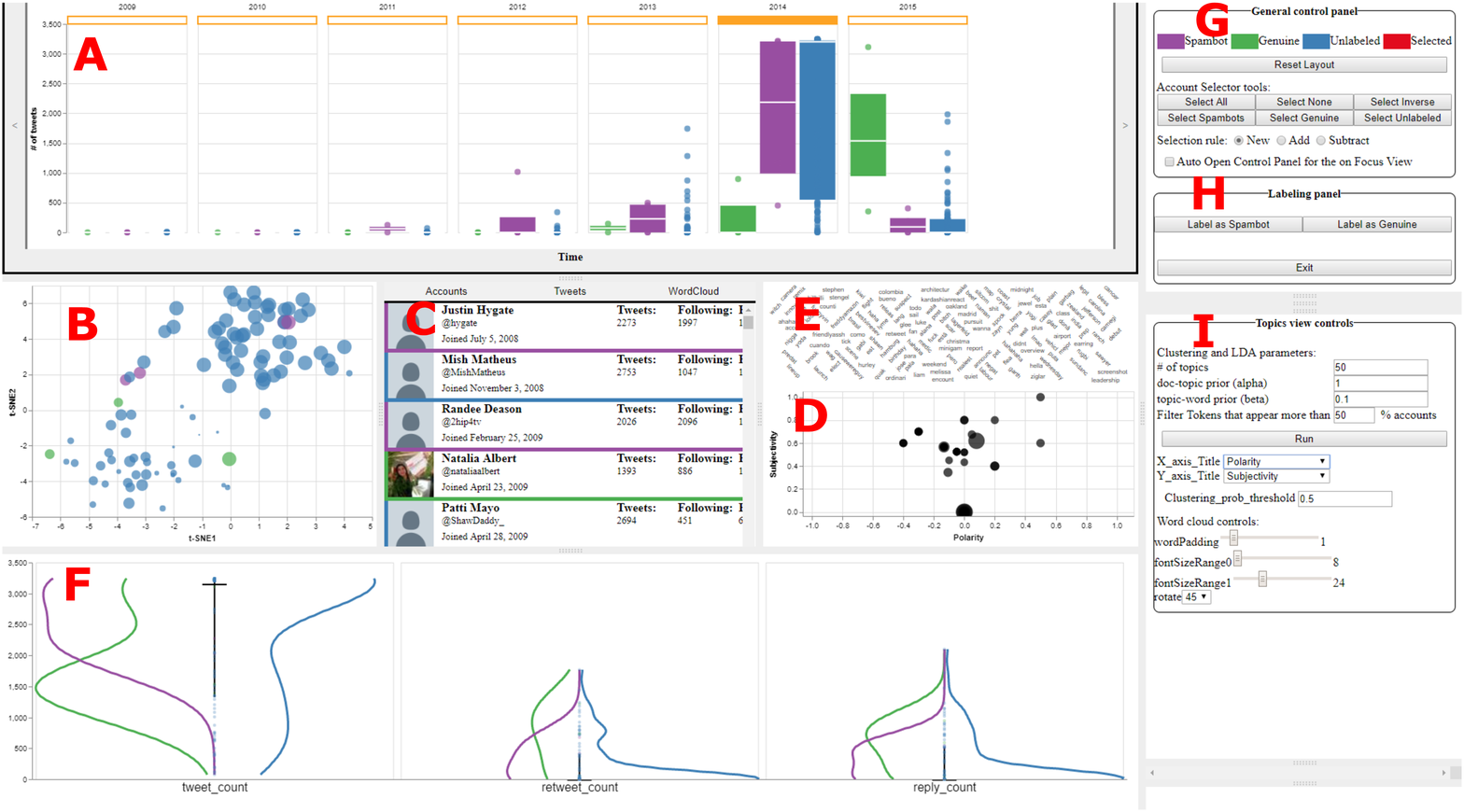}
  \caption[foo bar]{The default layout of the front-end of VASSL:
\begin {enumerate*}[label=\Alph*\upshape)]
\item the timeline view,
\item the dimensionality reduction view, 
\item the user/tweet detail views,
\item \& \item the topics view (clustering / words),
\item the feature explorer view,
\item the general control panel,
\item the labeling panel, and
\item the control panels for all the views (the opened control panel in the figure is for topics clustering view).
\end{enumerate*}}
	\label{fig:main}
}

\vgtcinsertpkg

\begin{document}


\firstsection{Introduction}
\maketitle

A social spambot is a computer algorithm that automatically produces content and interacts with humans on social media, trying to emulate and possibly alter their behavior \cite{ferrara2016rise}. These types of bots have been used to propagate harmful content such as spreading radicalism \cite{subrahmanian2016darpa}. To counter this threat, many automatic solutions have been designed to detect spambot accounts. However, the nature of the problem requires the continuous tracking of the performance of these models. As in many cybersecurity topics, malicious actors dynamically change and evade existing solutions. Researchers report a shift in the behavior of social spambots, which allows them to evade current solutions \cite{cresci2017paradigm}. Detecting this new generation of spambots individually has become more challenging at the account-level since they tend to propagate spam through campaigns rather than using single accounts \cite{zhang2016rise}. These accounts' behavior may be similar to genuine users in every aspect if monitored at the individual account-level. It has been shown that detecting this type of spambot with accepted accuracy is significantly more challenging for both human annotators and machine learning models \cite{cresci2017paradigm}; as a result, there is a need for new detection methods that are scalable and capable of tackling the dynamically changing environment. To address this challenge, new solutions have emerged to provide analysis of spambots at a group-level \cite{viswanath2015strength,cresci2018social}. Although it has been shown that these solutions significantly improve detection of the new generation of spambots in some instances, they are still unable to achieve the desired performance in general.

To tackle the problem of detecting this new type of spambot, we present VASSL, a visual analytics system that expedites and facilitates the process of spambot labeling. VASSL leverages multiple integrated computational and visual features to support human annotators in inspecting accounts from different angles and at different aggregation levels. Notably, it enables behaviour analysis of multiple accounts as groups, instead of analyzing accounts individually, enabling the detection of spammers using multiple accounts, as well as providing users with insights into the collective and dynamic behavior of spambots. VASSL also allows users to conduct analyses at a lower resolution, using views that reveal detailed information about a selected account.

VASSL is designed to work with Twitter accounts; however, the general concepts can be used with other social media platforms as well. VASSL provides five integrated interactive views that communicate different information about the accounts to support the process of labeling and is designed for use by analysts or expert users whose goal is to  efficiently and effectively annotate spambots and/or gain insight about online spambot behavior.

After presenting the design, algorithms, and various components of VASSL, we provide a use case that demonstrates the utilization of the developed tool to perform a complete analysis and labeling tasks. The benchmark dataset used in the testing was crawled from Twitter and prepared by \cite{cresci2017paradigm}, who published the data for research purposes.

To validate the usefulness of VASSL, we present a formal user study that compares our tool's labeling performance against a manual labeling approach.
The results indicate statistically significant improvement in the performance of human annotators when they use VASSL. VASSL improves the average accuracy of labeling by 14.6\% while increasing the effectiveness of labeling by 0.84 account per minute.

VASSL is publicly available at https://vassl.new-dimension.co. Besides incorporating existing visualization techniques, we present and integrate two novel interactive visualizations to communicate patterns of groups in time series data and to communicate the distributions of groups in multi-dimensional feature space.

This paper is organized as follows: Section 2 presents related work. In section 3, we present an overview of the design of VASSL and the requirements we seek to satisfy. Section 4 describes the functionalities of the back-end and provides more details about the techniques we use to prepare the data for the front-end, which is discussed in section 5. We present a use case of VASSL in section 6, followed by a description of our user study, which is presented in section 7. We discuss the results of the user study in section 8 and conclude the paper in section 9 with directions for future work.

\section{Related Work} \label{sec:relatedWork}
Previous research related to our focus in this paper can broadly be classified into two main categories: automatic spambot detection solutions and visual social media analytics.

\subsection{Automatic spambot detection solutions}
The problem of detecting spambots in social media applications such as Twitter is an active area of research \cite{jin2013understanding}. Researchers have proposed automatic algorithms to tackle this problem. These studies mainly differ in 
\begin {enumerate*}[label=\itshape\alph*\upshape)]
\item the process of generating labeled data which is used in training or testing, 
\item the features that are extracted and engineered to distinguish spambots from genuine accounts, and
\item the models that are used for the detection process.
\end{enumerate*}
 
Ground-truth, annotated (labeled) data to experiment is essential for designing and evaluating spambot detection methods. For this purpose, researchers use different methods to generate labeled data. As in many machine learning contexts, researchers can manually tag social media accounts by carefully inspecting the information available about them \cite{chu2010tweeting}. This method is not scalable, which may force researchers to rely on crowdsourcing solutions to generate the ground truth \cite{dickerson2014using}. However, achieving an acceptable level of reliability through crowdsourcing can be costly \cite{ferrara2016rise}, especially if the annotation tools are not efficient. Another method is to rely on sites' suspension mechanisms and crawl accounts that have been suspended by the sites' administrators \cite{morstatter2016new}. This method may not be accurate since it is possible for site administrators to ban accounts that are not spambots, e.g. for violating the site's rules. The last common method for building labeled data is by using what is known as a social honey-pot \cite{lee2011seven,stringhini2010detecting}. In this method, researchers create inactive accounts, the honey-pots, which do not interact with social media users or initiate any attractive posts. These characteristics reduce the probability of attracting genuine users and allow researchers to inspect accounts that interact with the honey-pots and possibly tag them as spambots.

Studies that propose automatic detection solutions commonly start with feature extraction and engineering. In this step, researchers define a set of features that allow automated detection models to separate genuine users from spambots. A rich set of features has been suggested in the literature and shown to be useful for the detection tasks. These features can be extracted from accounts' metadata, accounts' social networks, tweets' content, activity in time, sentiment, etc. \cite{varol2017online,dickerson2014using,morstatter2016new}.

After extracting this set of features, researchers propose different models that automate the detection of spambots. Most of these models are developed according to machine learning approaches, which can be categorized into supervised and unsupervised models. Supervised solutions utilize a set of labeled data to learn a discriminant function defined in the feature space. This function defines a boundary between spambots and genuine accounts, which is used afterwards to classify accounts according to their place in the defined feature space \cite{davis2016botornot, yang2013empirical}. On the other hand, unsupervised solutions build boundaries in the feature space according to similarities among accounts without utilizing predefined account labels \cite{miller2014Twitter, ahmed2013generic}. These boundaries define clusters of accounts that are considered similar, thus helping separate spambots from genuine accounts. Unsupervised models need additional human efforts to accurately label members of clusters. This requires an understanding of the characteristics of clusters produced by exploring their members, which may not be intuitive. Our visual analytics approach differs from these solutions, as it incorporates a human into the process of generating the initial clusters.

\subsection{Visual social media analytics}
The second group of related work is devoted to analyzing social media using visual analytics solutions. These studies may not target the spambot detection problem directly, but the visual analytics approach to analyzing social media has similarities to our approach. Moreover, many of the techniques proposed in these studies can be adopted for spambot detection.

Social media sites are rich sources of information that enable researchers to perform different kinds of analysis, including event analysis and cross-platform information linking \cite{chae2012spatiotemporal, andrienko2013thematic, chen2018sequence, lu2018visual}, visual sentiment analysis \cite{kucher2018state}, opinion diffusion analysis \cite{wu2014opinionflow, chen2016d,zhao2014fluxflow}, and text analysis and topic modeling \cite{makki2018atr,el2018progressive}. A thorough review of visual analytics research using social media data is presented in \cite{chen2017social,wu2016survey}.

The most relevant work to ours was conducted by Cao et. al. \cite{cao2016targetvue}, who proposed TargetVue, a visual analytics solution for detecting and analyzing anomalous user behavior in social media. The computation module in TargetVue utilizes the Time-adaptive Local Outlier Factor \enquote{TLOF} model to rank accounts according to their abnormal behavior. These accounts are visualized using multiple interactive views that allow the exploration of accounts and facilitate manual labeling.  One shortcoming of this work appears in the process of identifying the new type of spambot, which propagates malicious content using multiple accounts serving a single actor, rather than the older individual spambot accounts. In such cases, anomaly detection models working on the account-level, such as TLOF, are less effective in identifying spambots, as they could potentially create a \textit{rare category} \cite{lin2018rclens}. Our work emphasizes visualizing algorithmically derived information that helps in grouping accounts, enabling users to search for visual patterns that highlight malicious clusters rather than individual malicious accounts.

\section{Design Requirements}
To derive the initial design of VASSL, we consulted the literature on both social spambot labeling and social media visual analytics (covered in section \ref{sec:relatedWork}), with specific attention to the representation of accounts, potential automatic and interactive clustering methods, and visualization of high dimensional and textual data. After reviewing related literature, we derived the following set of system requirements:
\begin{itemize}
\item[\textbf{R1}] \textbf{Show similarities among accounts.} This requirement is essential for enabling users to explore different characteristics to cluster the accounts based on \cite{cresci2018social,cresci2017paradigm,cao2016targetvue}.
\item[\textbf{R2}] \textbf{Represent accounts at different aggregation levels.} Most of the features we found in the spambot detection literature can be considered time-series features. Examining these features at different temporal aggregation levels reveals different patterns which could help identify spambots \cite{cao2016targetvue}.
\item[\textbf{R3}] \textbf{Summarize tweets' content and show details on demand.} According to Shniderman's visual information-seeking mantra \cite{shneiderman2003eyes}, it is desirable to visualize the content summary of tweets to the users and show the details of the tweets as interactively requested.
\item[\textbf{R4}] \textbf{Allow the user to highlight and analyze groups of accounts.} The system should enable users to highlight and cluster accounts interactively.  This interactive clustering is essential to reveal potential group-based spamming activities \cite{cresci2018social,cresci2017paradigm}.
\item[\textbf{R5}] \textbf{Enhance the efficiency and effectiveness of human workers.} The system should enable cost effective annotation and improve the accuracy of detecting spambots as well as reducing the time needed to label groups of accounts by human workers \cite{ferrara2016rise}.
\item[\textbf{R6}] \textbf{Support different machines and screen sizes.} Most of the labeling tasks are performed on the web. A system that is intended to work in this environment should support workers with different operating systems and limited processing capabilities, and should adapt to various screen sizes\cite{leavitt2006based}.
\end{itemize}

With these requirements, we designed VASSL as a web-based application. The design assigns most of the computation intensive processes to the back-end of the system to reduce the load on the clients' machines. Moreover, the front-end is designed to have a flexible, responsive layout that allows the user to control the size of different views interactively. This feature is useful in focusing users' attention to particular parts of the information provided by VASSL. It is also helpful in generating a proper layout for different screen sizes for different clients. Both of these features were designed to satisfy (\textbf{R6}).

The most influential requirement that guides our system design is (\textbf{R1}). We built the entire system with the idea of showing similarities among accounts from different angles to support the interactive, human-guided clustering of accounts (\textbf{R4}). Four of the views in Fig.\ref{fig:main} are capable of showing similarities among the accounts. The timeline view, in particular, is designed to show similarities among accounts using different temporal aggregation levels to satisfy (\textbf{R2}). Once a potential cluster of accounts is found, different interactive selection methods of VASSL can be used to highlight  groups of accounts (\textbf{R4}), examine them in more detail (\textbf{R3}) and simultaneously label them as spambot or genuine. Such analysis saves human workers significant time (\textbf{R5}) because it allows the examination and labeling of accounts in parallel (Section \ref{sec:evaluation}), unlike the traditional manual labeling methods, which sequentially analyze and tag accounts one by one. Furthermore, the proposed parallel analysis enhances the effectiveness of human workers (\textbf{R5}) by revealing valuable insights needed to detect the new type of spambot that spreads spam in groups (Section \ref{sec:evaluation}).

To summarize the content of tweets (\textbf{R3}), we included two techniques in the system design; word cloud visualization \cite{seifert2008beauty} and topic modeling using Latent Dirichlet Allocation (LDA) \cite{blei2003latent}. The former is a well-known text visualization technique which reveals words that frequently appear in accounts tweets. Topic modeling, on the other hand, was chosen to show similarities among accounts in terms of their posting topics, which are used as a way to cluster accounts.

\subsection{Targeted Users}\label{sec:targetedUsers}
Our system utilizes sophisticated techniques and visualizations that require training and expertise. Our goal is to provide new functionalities to more effectively and efficiently identify spambots,  to ultimately reduce the time and cost of recruiting expert annotators and the annotation process (\textbf{R5}). The targeted users are human annotators whose terminal goal might be generating labelled datasets, or using the tool to understand and characterize spambot behaviour in a dataset. This guided us through many design choices with a focus on utility. However, for these functionalities to be useful, the users are required to have a certain level of expertise.

We made several assumptions about user expertise to operate our system effectively. The most important was users' knowledge of tuning machine learning models; specifically, dimensionality reduction and topic modeling. VASSL is designed to provide experts with interactive control of these techniques which rely heavily on parameters tuning. Familiarity with these tools and experience labeling social spambots will significantly improve users' experience with VASSL.

\section{Data processing and Analysis} \label{sec:Back_End}
In this section, we describe the main functionalities of the back-end of VASSL, which can be summarized as:
\begin {enumerate*}[label=\itshape\alph*\upshape)]
\item feature extraction,
\item dimensionality reduction, 
\item topic modeling, and
\item data communication.
\end{enumerate*}

The first functionality is the extraction of a set of features that represent Twitter accounts. We built on previous research to check the types of features that are known to be useful in spambot detection  \cite{varol2017online,dickerson2014using,morstatter2016new}. We identified a set of fifty features, e.g. total tweet count, average number of links, followers to following ratio, then generated four representations of these features by aggregating them temporally (\textbf{R2}). All extracted features are listed in the supplementary materials.

Some of the defined features are derived by applying sentiment analysis to accounts' tweets, which is calculated offline. These features are useful for identifying spambot accounts \cite{varol2017online} and can also provide overview information about the content of tweets (\textbf{R3}). We extracted several features that reflect the sentiment of accounts' tweets such as polarity and subjectivity scores. These scores are calculated by matching tweet words to corpus words that have been pre-labeled, e.g. movie review corpus \cite{Pang+Lee:04a}. We tokenize tweets' text, assign scores to the tokens by matching them with the labeled corpus words, then calculate the sentiment by taking a weighted sum of the assigned tokens' scores. Similarly to other features, the sentiment polarity and subjectivity is aggregated for each account. This aggregation generates time-independent features (aggregation over all tweets of the account) and three sets of time-dependent features (aggregation by year, month, or day). Because of the size of the data, this analysis is conducted offline as a preprocessing step.

The second functionality performed by the back-end is the generation of a two-dimensional representation of the accounts, appropriate for clustering tweets in a 2D display (\textbf{R1}). This reduction is useful to communicate similarities between accounts to the user. VASSL uses the extracted time-independent features for this purpose. We incorporate four different dimensionality reduction (DR) techniques: Kernel Principle component analysis (K-PCA) \cite{scholkopf1999kernel}, Linear Discriminant Analysis (LDA) \cite{mclachlan2004discriminant}, Locally Linear Embedding (LLE) \cite{roweis2000nonlinear}, and t-distributed Stochastic Neighbor Embedding (t-SNE) \cite{maaten2008visualizing}. The four DR techniques are included in order to increase the effectiveness of the dimensionality reduction results in different contexts. For example, if the task is to label a set of unlabeled data without any information about labeled data, supervised DR methods such as LDA may not produce good results, unlike PCA or LLE, which are unsupervised solutions. However, if a user has already labeled parts of the data, she may utilize a supervised DR method for better class separation performance. Another factor that encourages our choice of DR techniques is the assumption of linearity. We give users multiple options for reducing the dimension of feature space using linear and non-linear mapping techniques.

VASSL supports two transformation methods which can be incorporated with the aforementioned dimensionality reduction methods: min-max normalization and standardization. Min-max normalization changes the range of the features and forces it to the range from 0 to 1. Standardization transforms the values to Z-scores. Such transformations are needed for some of the dimensionality reduction techniques. For example, PCA is known to be sensitive to differences in features variance and thus may performs badly if applied before normalization.

The third functionality of the back-end is topic modeling, which is performed by utilizing the Latent Dirichlet Allocation (LDA) model\cite{blei2003latent}. We use the generated topics as a way to cluster accounts (\textbf{R1}, \textbf{R4}), as explained in Section \ref{sec:FrontEnd_TopicClustering}. VASSL employs multiple natural language processing techniques, such as lemmatization and stemming preprocessing, to transform the set of tweets for each account to tokens in a form suitable for LDA. The system then applies LDA to the set of tokens and generates a set of topics that best represent accounts' tweets. To increase the accuracy of the latent topics, the system applies LDA to each temporal aggregation level mentioned above.

One of the challenges we faced during VASSL development was the data scale. To discover anomalous botnet spammers, analyzing large numbers of accounts simultaneously is an ideal way to reveal patterns. However, the size of the data handled increases exponentially with the number of accounts, because we need to consider multiple representations of each account. To overcome this issue, the back-end keeps a communication channel open with connected clients, gradually feeding data. This channel is a querying mechanism between the front-end and the back-end, which only sends data that is visible in the views of the front-end and prepares the remaining in the back-end. This reduces the problem of information overload and creates a better analysis experience for the users.

The communication channel also facilitates modification of the behavior of automatic data analysis techniques according to users' input. For example, users are able to change the parameters of dimensionality reduction and topic modeling techniques from the front-end.

\section{Visualization and Interaction Design}

The front-end of VASSL consists of five views along with their control panels (Fig. \ref{fig:main}). 

\subsection{Timeline View}

The timeline view visualizes the distribution of time series features that represent Twitter accounts at three different aggregation levels. Influenced by \cite{williamson1989box}, we choose to use the box plot for its simplicity yet capability of visualizing complex multivariate data, such as our time series. The design of the visualization combines a bubble chart and a box plot to enable the user to select individual accounts, while observing class statistics (\textbf{R1}). Accounts are visualized in this view as points in temporally sorted facets. Each account has a representation in each facet to communicate changes in time. The accounts are grouped in a facet according to assigned labels into genuine, spambot, or unlabeled, which are the three levels of the x-axis of the facet. These groups are color-coded as green, purple and blue respectively. The y-axis of the timeline view represents one or more time series such as the total tweet count and the average number of hashtags in tweets, depending on user selection. The orange boxes on top of each facet are temporal selectors which can be used in temporal zooming interaction as explained below.

VASSL supports three main user interactions with the timeline view. Hovering over the quartile boxes increases their transparency, which helps users examine the underlying distribution of accounts underneath the boxes. Users can also zoom in time by moving the mouse pointer inside a facet and scrolling up and down to zoom in and out. Zooming functionality changes the aggregation level of the time series to year, month, or day levels (\textbf{R2}, \textbf{R3}).

The last interaction supported in the timeline view is the account selection. Selected accounts are highlighted using red color. Users can click on the points representing the accounts to select and highlight the accounts. Users can also select accounts by brushing on any facet to select accounts that overlap with the brush. VASSL highlights selected accounts in every time facet as well as in all other views.  Linking time series allows users to examine trends and anomalies for selected accounts over time. Moreover, linking the views allows users to examine different information about the selected accounts such as their tweets, their position in the feature space, etc (\textbf{R4}).

\begin{figure}[t]
 \centering 
 \includegraphics[width=\columnwidth]{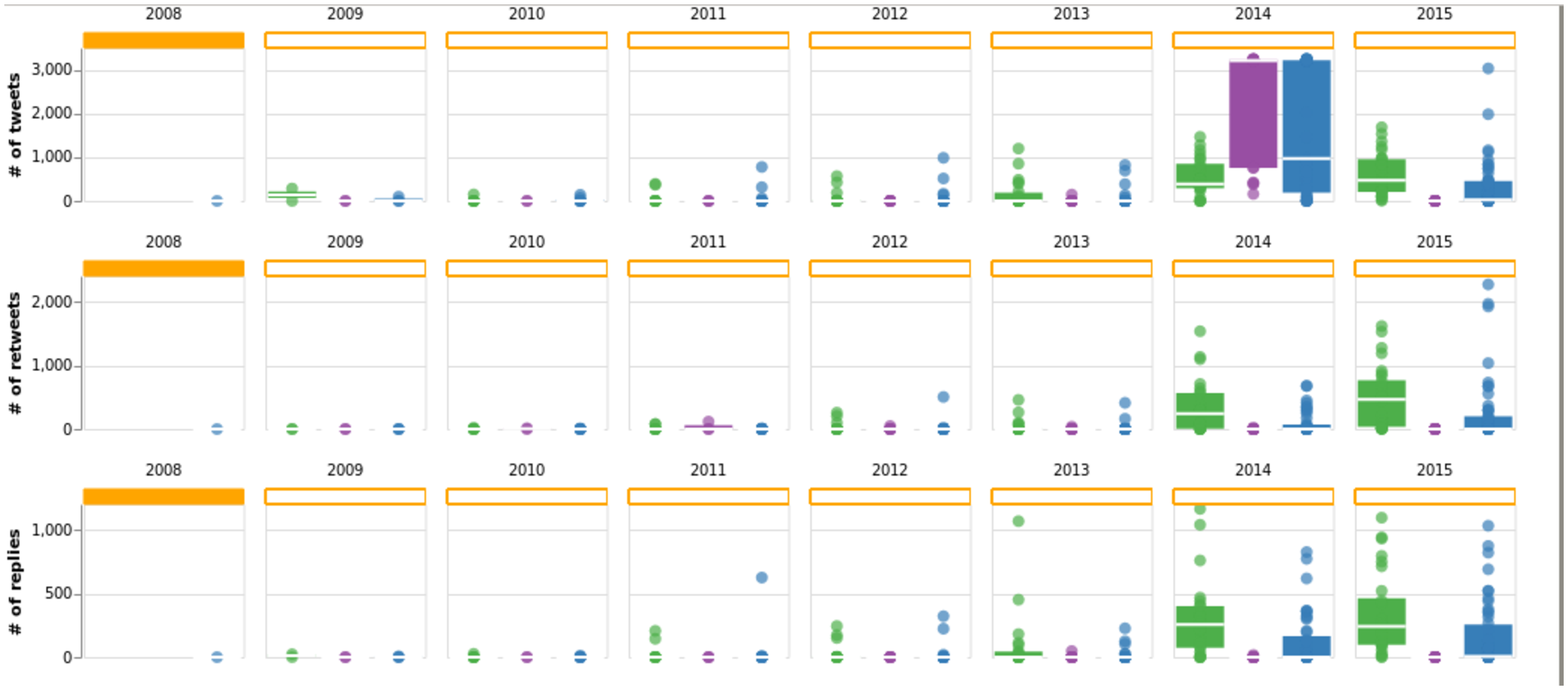}
 \caption{The timeline view visualizing three time series features at the year aggregation level. Blue shows unlabeled accounts, green for genuine accounts, and purple indicates spambots.}
 \label{fig:timeline}
\end{figure}

Users can view multiple time series, which  divides the visual space of the timeline view among selected time series (Fig. \ref{fig:timeline}). This allows simultaneous exploration of multiple time series features using a small-multiple-like representation \cite{van2013small}. The control panel also enables users to change the temporal resolution of the view. The visualizations of time series automatically adjust to match the selected time resolution.

\subsection{Dimensionality Reduction View} \label{sec:DR_view}

VASSL visualizes the results of the dimensionality reduction (DR) techniques, explained in Section \ref{sec:Back_End}, in a 2D scatterplot (Fig. \ref{fig:main} (B)). The effectiveness of 2D scatterplots in visualizing DR results, while maintaining cluster separability has been shown in \cite{sedlmair2013empirical}. Our scatterplot allows exploring similarities among accounts (\textbf{R1}), which are color-coded according to their class. The bubble size  represents the tweet count feature of each account. Similar to the timeline view, users can select accounts by clicking or brushing. Moreover, the view can be panned and zoomed as needed.

The users can select the specific DR technique to be used in the view from the four supported DR techniques. Users can also tune the hyperparameters of the techniques. For example, users can select the kernel function to be used with PCA, the number of neighbors to be considered in LLE, the perplexity of t-SNE, etc. The control panel also allows users to select a pre-reduction transformation including min/max normalization and score standardization, as explained in Section \ref{sec:Back_End}.

\subsection{Users/Tweets Details Views}

Three tabs are used in the details view: accounts' cards view, tweets view, and tweets' word cloud view. The accounts' cards view shows a list of accounts and other useful information, including accounts' names/screen names, profile images, joining date, total tweets, number of followers and followees, and the number of likes. 

Users can select an account by clicking on its card (linked to other views). Once an account or a set of accounts are selected, users can access the tweets view, which shows the selected accounts' tweets in chronological order (\textbf{R3}). Including tweet text during the analysis is essential 
to utilize the human ability to detect automated text generation (as explained in Section \ref{sec:useCase}. Instead of accessing the entire tweet text, users can use the word cloud view to visualize selected tweets using the word cloud visualization technique \cite{seifert2008beauty}. The word cloud visualization is helpful for revealing repetitions of wording in selected tweets, which can guide the exploration of tweet text and labeling (\textbf{R3}).

\begin{figure}[!t]
 \centering 
 \includegraphics[width=\columnwidth]{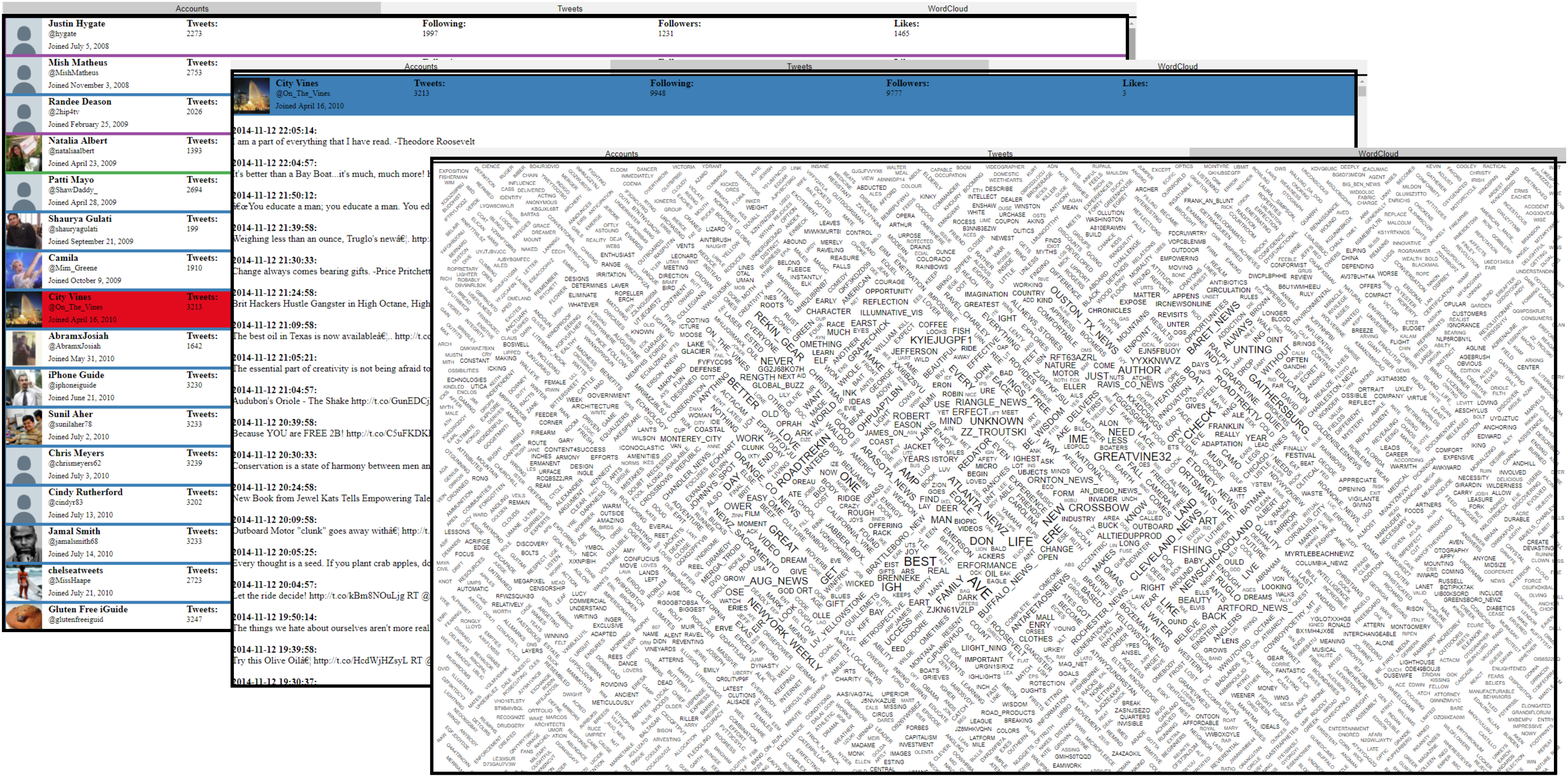}
 \caption{The three details views. Selecting one or more accounts from the cards view shows their tweets in the tweets view and a word cloud of all the tweets in the word cloud view.}
 \label{fig:detailsViews}
\end{figure}

\subsection{Topic Clustering View}\label{sec:FrontEnd_TopicClustering}

This topic clustering view uses two visualizations, which help analyze the results of applying Latent Dirichlet Allocation (LDA) to accounts' tweets. The first visualization is a bubble chart that represents the generated latent topics in a two-dimensional space (Fig. \ref{fig:main} (D)). The axes of this visualization can be chosen from the control panel to be either unique IDs, topics polarity, or topics subjectivity, which are calculated by applying sentiment analysis to the topics' most probable words. The size of the bubble encodes a score for each topic which represents the sum of probabilities of posting in that topic by all the accounts (see equation \ref{eq:topicScores}). The score of a topic $T_i$ is calculated by summing up the probability of that topic in all $j$ documents $D_n, \forall n \in \big\{1, 2, \dots, j\big\}$. Documents in our analysis are accounts represented by the concatenation of all their tweets.

\begin{equation}\label{eq:topicScores}
T_iScore = \sum_{n=1}^{j} P(T_i|D_n)
\end{equation}

The second visualization in the topic clustering view is a word cloud visualization of the most frequent words in generated topics. LDA computes probabilities that show the distribution of these words in each topic. VASSL utilizes these scores to determine the size of the words in the word cloud. When a user hovers over a topic, the word cloud changes the size of the words according to their relevance scores, which help the user in exploring the semantics of the topics.

\begin{figure}[th]
 \centering 
 \includegraphics[width=\columnwidth]{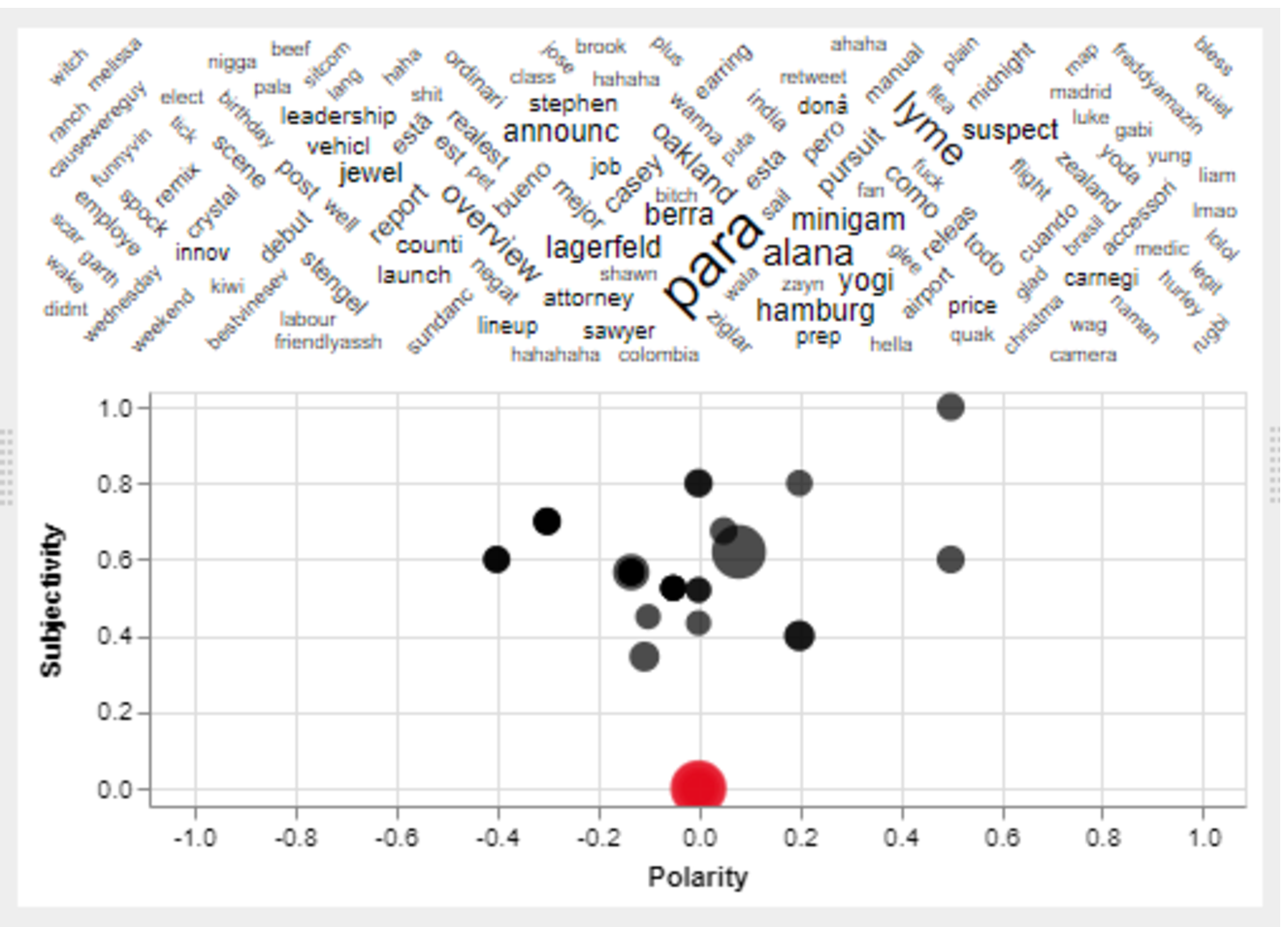}
 \caption{The topics clustering view with one topic selected. The bubble chart in the bottom shows the topics with bubble size communicating topics' scores. The word cloud on the top shows the topics' words with word sizes representing words-topics distribution.}
 \label{fig:topics}
\end{figure}

Users can interact with the topics bubble chart in multiple ways. Beside zooming and panning, the visualization supports selection and brushing of topics, which allows for exploration of each topic's words. When selecting multiple topics, VASSL aggregates words' probabilities to show their relevance to all selected topics (Fig.\ref{fig:topics}).

Selecting a topic results in the selection of all accounts that have posted in that topic, with probability more than a threshold selected by the user (\textbf{R1}, \textbf{R4}). In other words, we can consider the topics as clusters where a single user can belong to more than one cluster (fuzzy clustering). Changing the threshold controls the sensitivity of cluster membership. Increasing the threshold narrows the results to accounts that post in a topic frequently, while reducing the threshold includes accounts that may rarely post in a topic.

The control panel for topics clustering view enables users to tune LDA hyperparameters. This includes the number of latent topics to generate, and the alpha and beta parameters which control the distribution of documents/topics and topics/words respectively. VASSL uses symmetric Dirichlet distributions. Assigning low alpha value results in modeling an account's tweets with a few numbers of topics and vice versa. Similarly, using low beta values results in modeling a topic using few words and vice versa. Tuning the three parameters enables users to control the generated topics to increase the effectiveness of the LDA.

Topic modeling of tweets is performed with respect to the selected temporal resolution. The topic clustering view is linked to the timeline view. When a user zooms into a particular period using the timeline view, the topics clustering view updates the topics to match the selected period (\textbf{R2}). This enables users to examine the change of tweets topic in time. The trade-off of this flexibility lies in the efficiency of topic modeling; because of the massive size of the documents, the topic modeling procedure is not performed at interaction speed, producing lag times every time the hyperparameters are changed. However, by keeping the hyperparameters constant, the topic views are generated and cached in advance of interaction.

We evaluated the time needed to perform topic modeling of 100 accounts at year aggregation level (the worst case scenario) on a cloud server with Intel Xeon CPU E5-2650L v3 @1.80GHz. It took an average of 69.09 seconds (with 1.33 SD) for 15 trials with different hyperparameters to complete the topic modeling and visualize the results. We observe a strong correlation between completion time and topic count as anticipated, and we report the completion time according to a randomly generated number of topics in the range from 1 to 100.

To overcome the time limitation of topic modeling, we implemented two solutions in VASSL. The first is to generate LDA topics for predefined profiles (pre-determined assignment of the hyperparameters) during the preprocessing stage. This allows VASSL to load the topics for these profiles at interaction speed. This method, however, would only be useful in cases where users do not consider tuning the hyperparameters online. The second solution is the usage of a spinner wheel that only disables interaction with the topics clustering view during LDA back-end calculations. This allows the user to interact with all other views of VASSL while computation is completed.

\subsection{Feature Explorer View}

\begin{figure}[t]
 \centering 
 \includegraphics[width=\columnwidth]{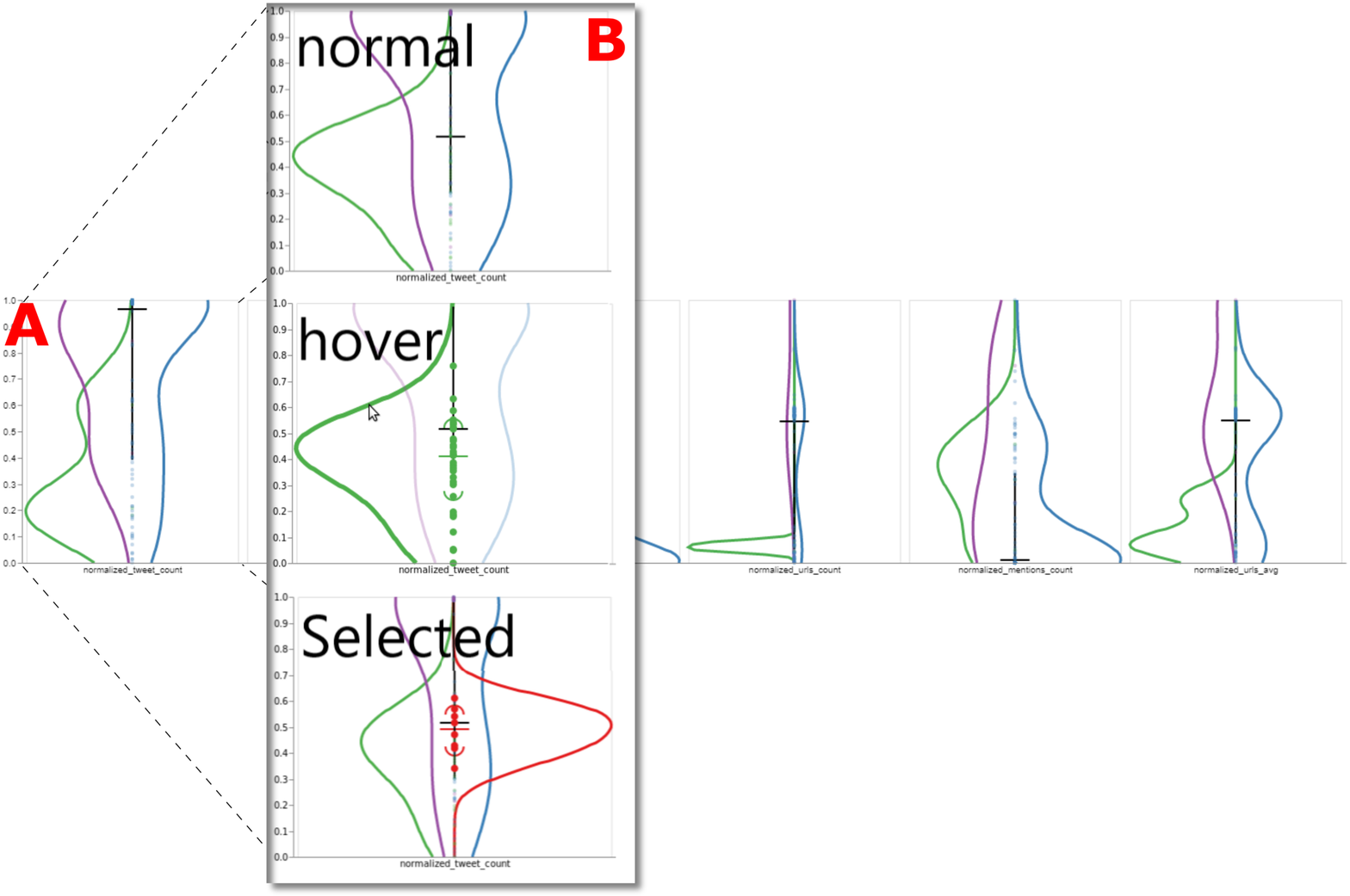}
 \caption{The Feature explorer view using a modified violin plot. The lines represent a kernel density estimation of classes PDFs. The accounts are distributed in the middle of each facet to facilitate selection. Blue represent unlabeled accounts, green for genuine and purple for spambots. A) Multiple facets representing a set of selected features. B) The effect of some user interaction on one of the selected features.}
 \label{fig:features}
\end{figure}

The Feature Explorer view visualizes the distribution of accounts in selected features using a new design based on a violin plot (see Fig. \ref{fig:features}). We used a violin plot instead of a box plot to enable the user to examine multi-modality in any feature \cite{hintze1998violin}, which could indicate a potential cluster (\textbf{R1}, \textbf{R4}). Users can select as many features as needed in the feature explorer control panel, which contains a list of all features (see Fig. \ref{fig:controlPanels}). The maximum number of features that can be visualized simultaneously depends on screen size and human perception. Selecting features divides the available visual space among the features, and thus can reduce the capability of absorbing communicated information.

The feature explorer view shows statistical summaries for each feature independently, such as median and quartiles of the overall accounts as well as labeled groups. Features are represented as multiple horizontally adjacent facets, having the same Y-axis range in order to correlate the features. The accounts are represented as points in the horizontal center of each facet. The vertical locations of the accounts in a facet are determined by the value of the feature for these accounts. To reduce the visual clutter of the feature explorer view, the accounts point is transparent by default. Hovering over a class distribution increases the opacity of the accounts that belong to the hovered class. The black solid line in the horizontal center of each facet represents the 1st and 3rd quartile of all accounts regardless of their class while the black tick mark represents the median of all accounts. Besides its job of communicating quartile information of all accounts, the solid black line divides the facet into two areas. The area on the left of the solid line is used to indicate spambot and genuine class distributions (purple and green curves respectively) which are approximated by kernel density estimation (KDE) technique. The right area is used to communicate the KDE of the unlabeled accounts distribution as well as selected accounts distribution (blue and red curves respectively). These curves are visualized in each facet in a similar manner to communicate this statistical information in each feature (\textbf{R4}).

Users can interact with the feature explorer view to examine class statistics in each feature or to select a set of accounts. Hovering over a distribution curve reveals more information about the class it represents, including the median, the 1st and 3rd quartiles, and the accounts belonging to that class. The feature explorer view supports selecting and brushing, which is linked to all other views. When a user selects a set of accounts, the feature explorer draws the distribution curves of highlighted accounts in each opened feature facet. Fig. \ref{fig:features} shows an example of five selected features as well as an illustration of the supported interactions. The control panel of the features explorer view allows the user to transform the features with similar transformations as described in Section \ref{sec:DR_view}, using min-max normalization or standardization to Z-scores, which is useful for comparison and outliers detection.

\subsection{The Control Panels Area}

The general control panel contains some functionalities that are applied to all views. It provides a legend that explains the color code we chose after consulting ColorBrewer \cite{HarrowerMark2003CAOT}. The general control panel also facilitates multiple selection tools and rules that allow the user to perform complex selection queries by simple button clicks. This includes selecting all accounts, selecting inverse, selecting none and selecting all accounts of a particular class type. VASSL supports three selection rules: New, Add and Subtract. Choosing the “New” rule unselects previously selected accounts, then selects whatever a user selects. Choosing the “Add” rule keeps existing selected accounts, and add whatever a user selects to the selection set. Choosing the “Subtract” rule removes whatever a user selects from any existing selection.

\begin{figure*}[t]
 \centering 
 \includegraphics[width=\linewidth]{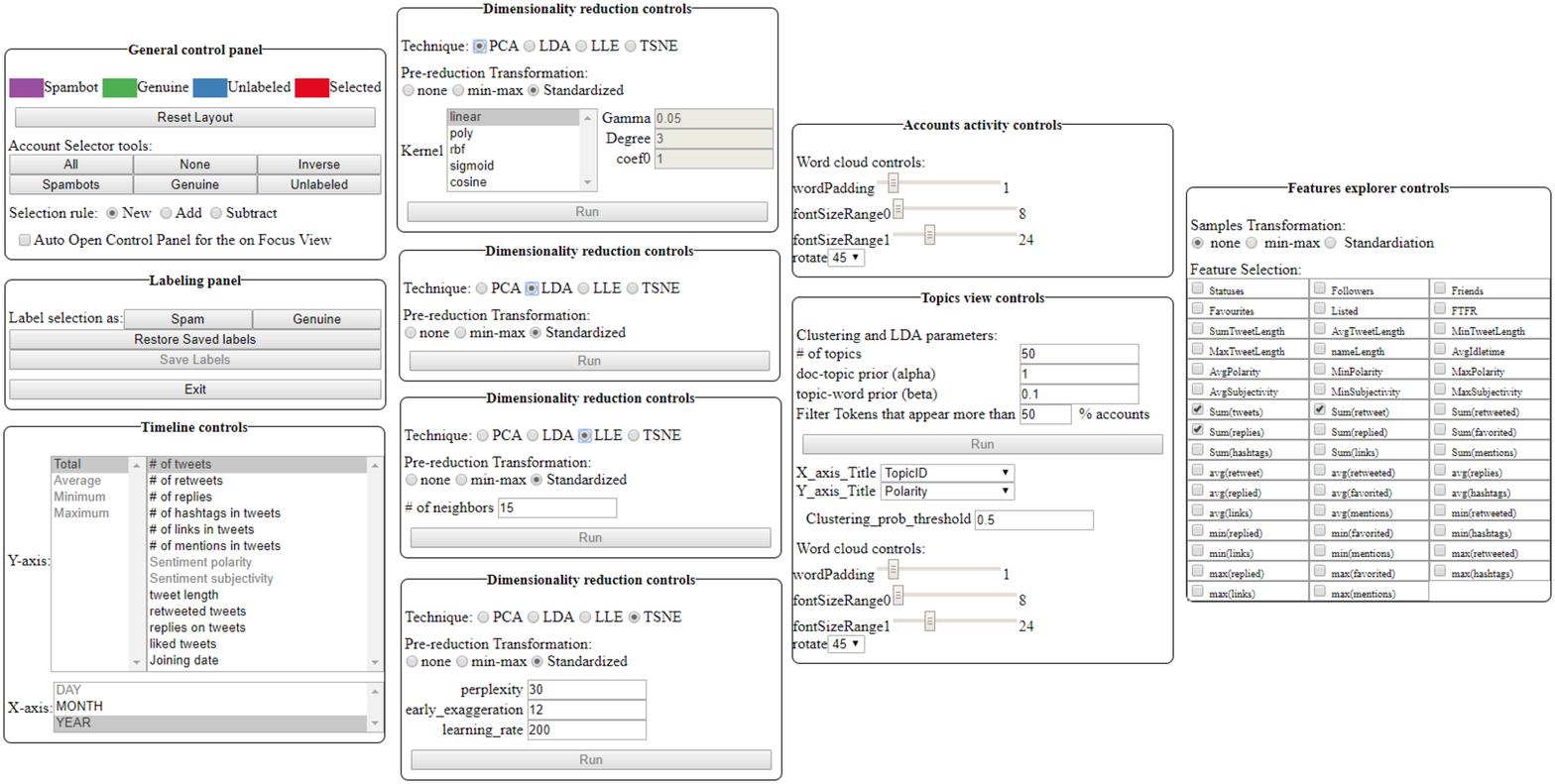}
 \caption{The control panels. Users can open a control panel of a view by pressing the \enquote{C} button on the keyboard while hovering over the view.}
 \label{fig:controlPanels}
\end{figure*}

Beneath the general control panel is the labeling panel, which allows the user to label selected accounts as spambots or genuine. Labels are automatically saved in a database managed by the back-end.
Fig. \ref{fig:controlPanels} shows the five specific control panels for each view.

\begin{figure*}[!h]
 \centering 
 \includegraphics[width=\linewidth]{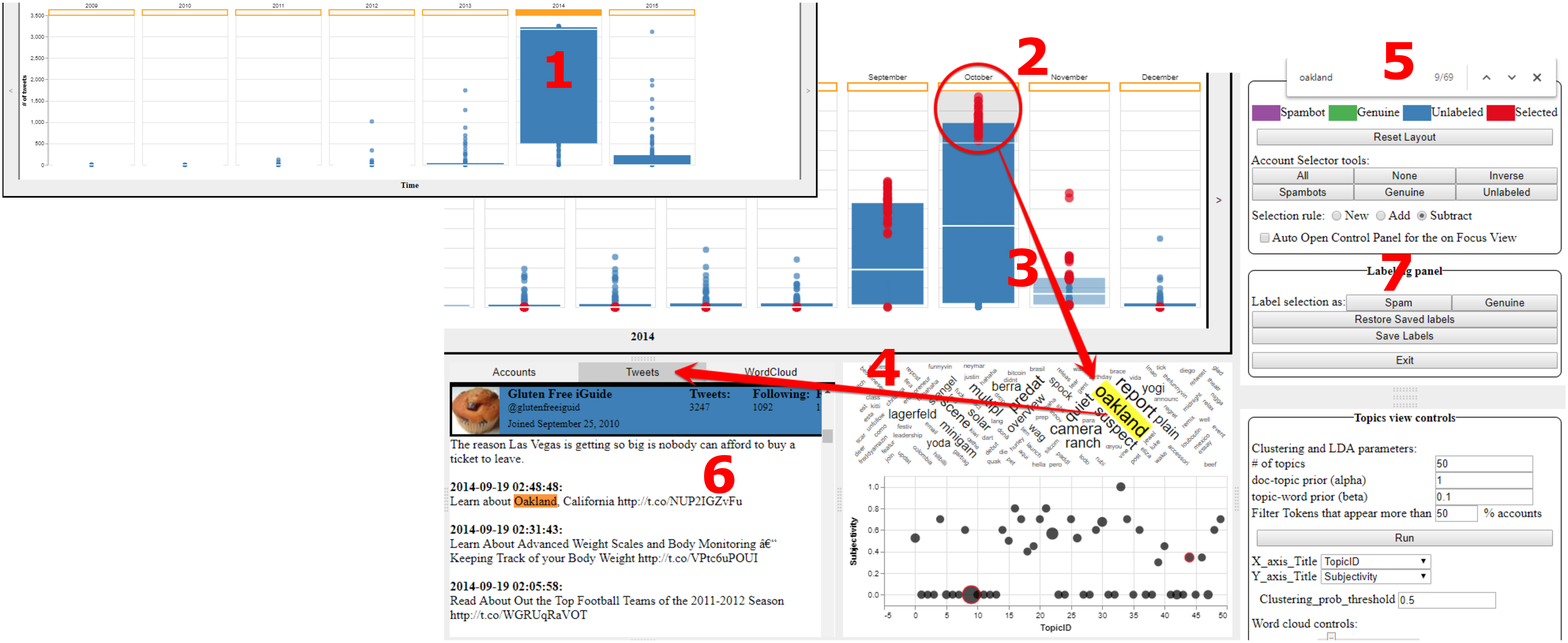}
 \caption{1) Abnormality in tweet count on 2014. Scrolling changes the timeline view to month-aggregation-level for the 2014 period. 2) Brush on the month of October 2014 to select a group of accounts with abnormal tweeting frequency. 3) Identify frequent words in the most frequent topics the selected accounts post in. 4) Open tweet view to examine selected accounts tweets. 5) Search tweets for the captured most frequent word in outlined topics. 6) Examine the tweets contain that word which leads to a discovery of automated-like tweets. 7) Label selected accounts as spambots.} 
 \label{fig:useCase}
\end{figure*}

\section{Use Case} \label{sec:useCase}
In this section, we demonstrate how to utilize VASSL to support the process of detecting and labeling spambots. This scenario is supposed to represent labeling tasks that are traditionally performed manually by recruited human workers who inspect and annotate a set of Twitter accounts. We take the role of the human workers and present the insights reached with the help of VASSL while interacting with its views. We note that this analysis aims to showcase how to use VASSL and what possible insights could be derived with it; the next section also evaluates the toolkit formally through a user study.

A human worker, called Amy, started her analysis by loading 100 Twitter accounts and their tweets to VASSL. Her first glance on the timeline view (Fig. \ref{fig:useCase} (1)) allowed her to spot an abnormal posting behavior in the year 2014. She zoomed into that period and examined the tweet count of the accounts in month-aggregation-level. She noticed that the unusual posting count happened in September and October where two groups could be separated based on tweeting frequency (Fig. \ref{fig:useCase} (2)). Amy selected the group with a high posting count, which highlighted their representation in all the views. VASSL also outlined the topics posted by these accounts and the words that represent these topics. Amy found that the word with the highest score for the outlined topics was \enquote{Oakland} (Fig. \ref{fig:useCase} (3)). She then examined the tweet view and searched for the word \enquote{Oakland} in selected accounts' tweets (Fig. \ref{fig:useCase} (4\&5\&6)). She found that all these accounts propagated the same tweet at different times in the months of September, October, and November of 2014. That tweet contained a link to an external website called \enquote{hub pages} that directed to a page with unfound contents. Examining other tweets posted by selected accounts showed a propagation of similar tweet content and links, but for other cities. Based on these findings, Amy to labeled selected accounts as spambots. The insights that led Amy to label these accounts as spambots were only reached as a result of analyzing multiple accounts as a group, which is the primary focus of VASSL.

After labeling the first group of accounts, Amy examined the dimensionality reduction view to find similar accounts. She found suspicious accounts that were not yet labeled and were very similar to the spambot accounts she had previously labeled. Amy examined the tweets of these accounts and found another account that propagated the same tweet about \enquote{Oakland}, so she included it with the spambots class.

Amy returned to the year-aggregation-level in timeline view and noticed that all the accounts that she labeled as spambots were inactive in all periods except in 2014 when they became abnormally active. All spambots posted more than 3000 tweets in 2014 with a very small variance among the accounts. By selecting all the accounts that tweeted more than 3000 times in 2014, Amy found one account that was unlabeled and had the same temporal pattern of posting as the group of spambots. She selected that account and compared it with the spambot accounts in the multidimensional feature space using the feature explorer view. She found that this account fitted well in the estimated distribution of spambots in most of the features, including tweet count, retweet count, reply count, URLs count, etc. This convinced Amy to label the account as a spambot even though the account tweeted on different topics than the pre-labeled spambots.

During her analysis, Amy gained a new insight into the sentiment of the topics. The group of spambots she had labeled usually tweeted about topics mostly constituted of words with very low subjectivity scores. This insight was derived from visualizing topics' bubbles on the sentiment subjectivity axis (Fig. \ref{fig:useCase}). Moving from this finding, Amy selected all the topics that had low subjectivity scores, which highlighted the accounts that frequently tweeted about them. Removing spambots from the selected list showed that the remaining accounts tweeted on a variety of low subjective topics, unlike spambots, which commonly tweeted on the same low subjective topics. After examining the tweets of unlabeled accounts that tweeted on low subjective topics, Amy could not find any suspicious tweeting behavior. Moreover, these accounts did not follow the distributions of spambots in the feature explorer view; as a result, she did not label them as spambots.

In the topics clustering view, Amy found two topics with high tweeting scores, i.e. many accounts posting in those two topics. Both topics had a high subjectivity score. Examining topics' words showed that they consisted of curses and slang, which tend to appear naturally in humans' everyday expressions. Selecting those topics highlighted the accounts that posted in them frequently. Amy noticed from the dimensionality reduction view that the highlighted accounts were dissimilar to labeled spambots. She confirmed this dissimilarity by examining the distribution of these accounts in the feature explorer view and found that they could be separated from the distribution of spambots in many features. After reviewing the tweets from these accounts, Amy decided to label them as genuine.

Amy noticed that the separation between spambots and genuine accounts was significantly affected by the subjectivity of tweet topics. Thus, she decided to select all the topics that had a subjectivity score of more than 0.5 (subjectivity range = [0,1]) and examined their words. Most of these topics had the same word-topic distribution if the semantics of the words were considered. Selecting similar topics with high subjectivity scores highlighted four unlabeled accounts which clearly belonged to the distributions of genuine accounts.

With these insights, Amy was able to label 85\% of the accounts. To label the remaining accounts, she examined them one by one and checked to see if she could observe similarities between them and any of the labeled accounts. At this stage, as it became harder to justify the labeling decision based on similarities only, Amy thoroughly examined each account's tweets. She started the exploration of tweets in the word cloud view to get an overview of the most frequent words that appeared in the accounts' tweets. A useful view in this stage was the dimensionality reduction view with linear discriminant analysis (LDA), to find the best 2-dimensions that separate the classes. Amy also opened ten features that had a clear distinction between the distribution of spambots and genuine accounts in the feature explorer view and kept track of the accounts' position in the feature space with respect to class distributions. Considering all the information that is communicated by VASSL, Amy was able to label all the accounts.

We compared the tags assigned by us (while assuming the role of Amy)  to the available ground truth. We achieved an accuracy of 95\% with one false positive and four false negatives for the spambot class. We labeled all 100 accounts within 15 minutes. This result was not meant to replace formal user study (the focus of the next section); we only provide the result to show how the steps taken in this section led to acceptable results. Identifying relationships among accounts and observing similarities in their behavior was critical to detecting the new type of social spambots.

\section{Evaluation} \label{sec:evaluation}
To evaluate VASSL, we conducted a within-subjects user study with college students to evaluate the accuracy of labeling Twitter accounts using VASSL and to compare it with the typical manual labeling procedure. We also collected subjective opinion from the participants to capture the perceived usefulness and ease of use of our toolkit.

\paragraph{\textbf{Participants}} We recruited 12 college students (11 male and 1 female). The major of most participants was Electrical and Computer Engineering. We limited the participation pool to individuals with basic knowledge about Twitter. Our participants were asked to work for up to 90 minutes and were compensated with \$10 for participation, plus \$5 as a motivation bonus when achieving $\geq85\%$ labeling accuracy.

\paragraph{\textbf{Methodology}} Participants were asked to complete four sessions. In session (A), we asked the subject to manually tag 100 unlabeled Twitter accounts using a tool that provides typical information that can be found in Twitter, such as the tweets of the account, joining date, the total number of tweets, etc. The tool provided a list of accounts that need to be labeled and allowed the user to select an account from the list, examine its tweets and the account's information, and tag it (spambot/genuine). A screenshot of the interface is available in the supplementary material. Participants were allowed not to tag an account at all.  We gave the participants 30 minutes to complete that task but allowed them to finish before the time limit if they chose.

In session (B), we asked the participants to complete a 20-minute training session to learn about the different views of VASSL. The training started with a 10 minutes tutorial video, followed by up to 10 minutes in which we allowed the participants to explore VASSL's various functionalities.

After completing the training session, participants completed session (C) to test VASSL. Similarly to session (A), the participants were asked to label 100 unlabeled Twitter accounts in up to 30 minutes, this time using VASSL. Our study did not consider the learnability factor, so we allowed subjects to ask us questions during the session if they did not understand the information communicated by the views.

We controlled the order of taking the sessions to eliminate the confounders that could appear due to carryover effects across testing sessions. We randomly assigned participants to one of two groups, to identify who would use which labeling solution first. This element of the activity is essential for balancing out human factors and ensuring the validity of our results.

After completing the three sessions above, participants took a 10-minute exit questionnaire to communicate their personal opinions about the tools. The quantitative responses we collected from the survey follow the seven-point Likert scale proposed by Davis \cite{davis1989perceived} to evaluate the perceived usefulness and ease of use. We also collected qualitative feedback from the participants to highlight issues in the current design of VASSL and suggest future research.

\paragraph{\textbf{Hypothesis}} The null hypothesis in our objective experiment, which we aim to disprove, is: \enquote{The mean of labeling accuracy for manual labeling procedure is equal to the mean of labeling accuracy achieved by VASSL}.

\paragraph{\textbf{Data set}}
During this experiment, we used the benchmark test set \#2 by \cite{cresci2017paradigm}. This data set contains 928 Twitter accounts (50\% spambots and 50\% genuine) and 2,628,181 tweets. The ground truth labels are available for this dataset. We took nine samples, each containing 100 accounts unique to that sample along with all its tweets. Each participant manually labeled one sample and tagged another sample with VASSL.

\begin{figure}[th]
 \centering 
 \includegraphics[width=\columnwidth]{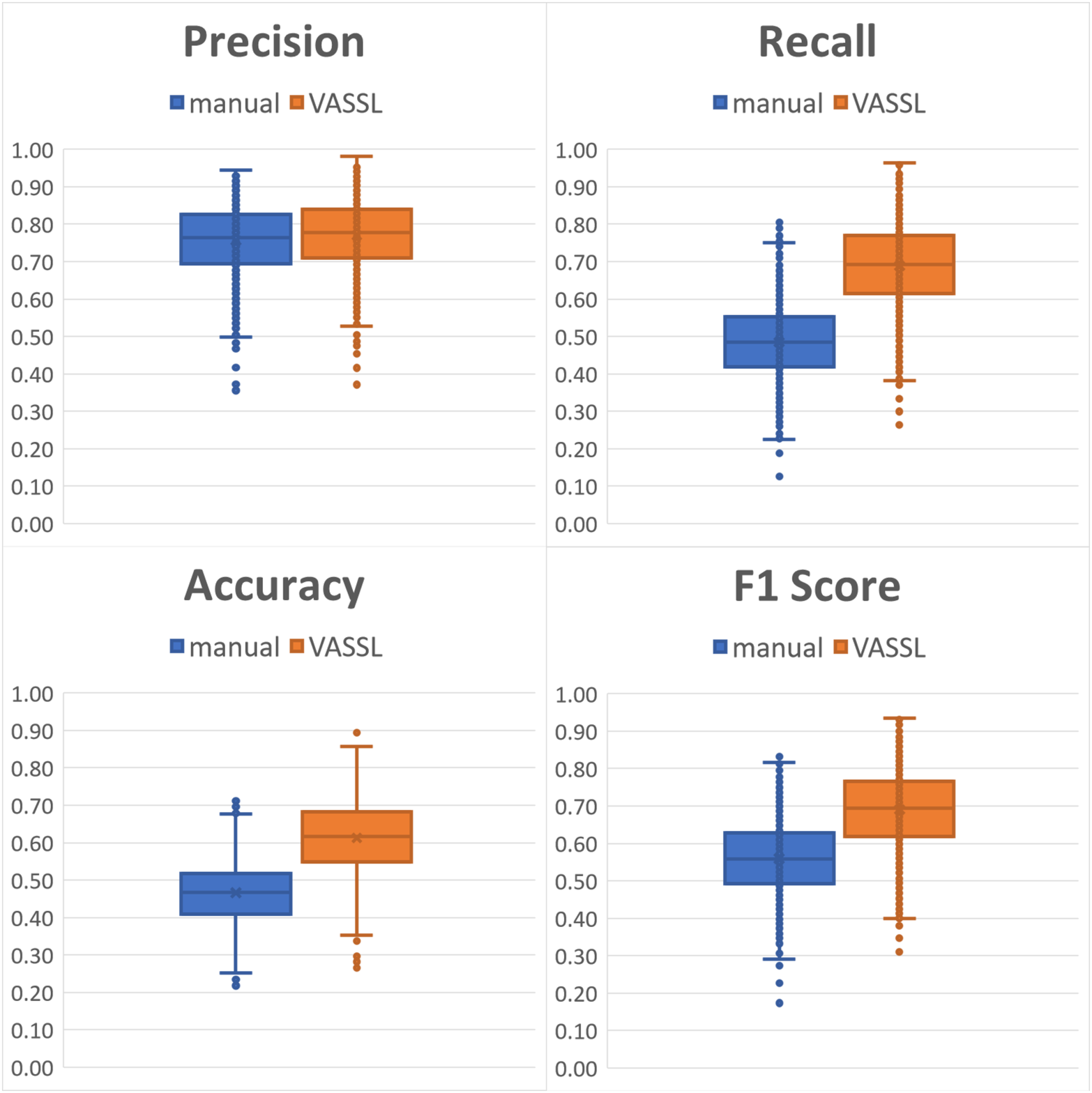}
 \caption{The objective performance of VASSL compared to manual labeling in terms of the precision, recall, and F1 score for the spambot class. We also include the overall accuracy of the labeling, which consider both spambot and genuine classes.}
 \label{fig:ObjectivePerformance}
\end{figure}

\begin{figure*}[th]
 \centering 
 \includegraphics[width=\linewidth]{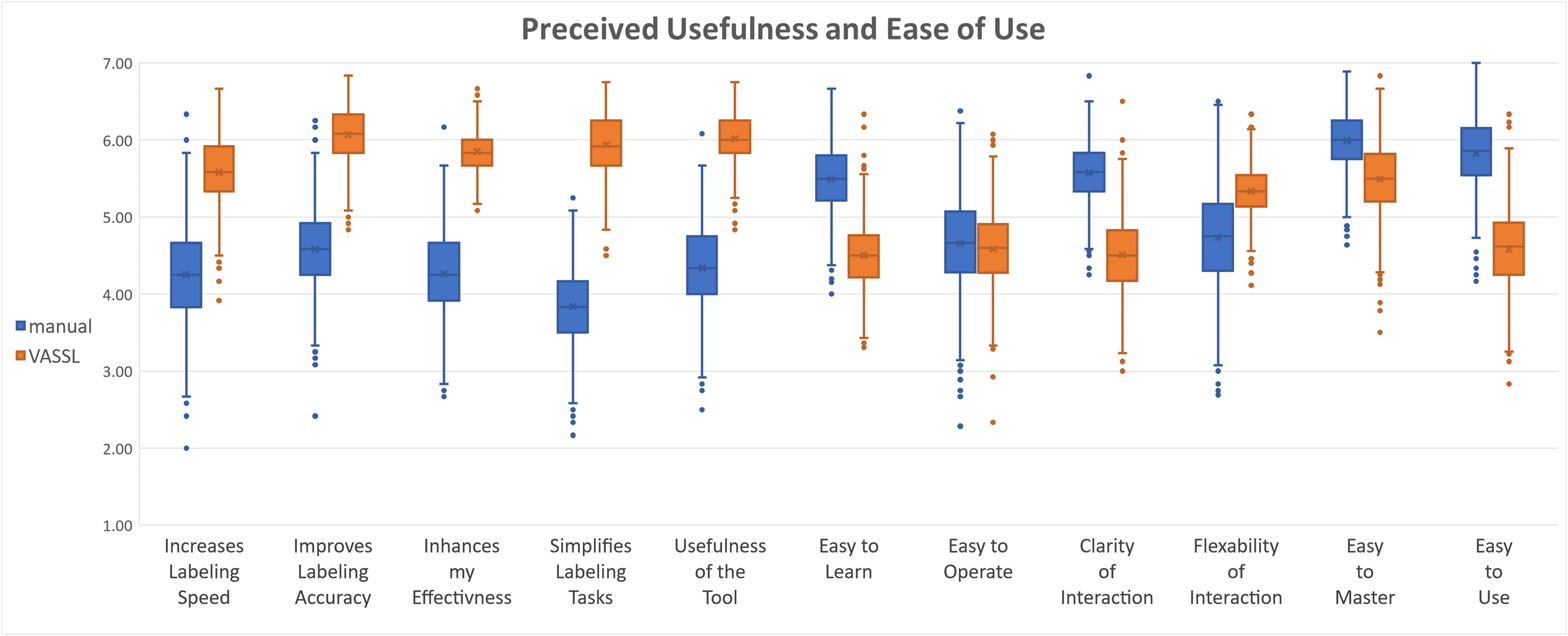}
 \caption{The subjective opinion about VASSL in terms of perceived usefulness and ease of use. The figure compares VASSL and manual labeling from user perspective.}
 \label{fig:SubjectiveOpinion}
\end{figure*}

\paragraph{\textbf{Results}}
Fig.\ref{fig:ObjectivePerformance} shows the performance of VASSL compared to manual labeling. We used four different metrics that measured the performance of labeling: precision, recall, accuracy, and F1 score. We bootstrapped the performance scores obtained from testing our participants and applied a single factor ANalysis Of VAriance (ANOVA) to test our null hypothesis for the four objective metrics. We found a significant improvement in the average accuracy [F(1,22)=9.7484, p=0.0049], average recall [F(1,22)=31.5232, p=0.00001], and average F1 [F(1,22)=13.6957, p=0.0012]. Adopting a significance level of 0.05, we reject the Null hypothesis for all these tests. However, we do not reject the Null hypothesis in the case of average precision [F(1,22)=0.8761, p=0.3594].

Another objective metric we collected to evaluate the effectiveness was the total number of tagged accounts in a test session. This metric replaced completion time, which could misrepresent participants who decided to end a session before tagging the entire test sample. Participants were able to label 76.36 accounts with VASSL on average, compared to 51.07 accounts for the manual labeling approach (with standard deviations of 7.9 and 8.92, respectively).

The bootstrapped results of the subjective scores we collected from the participants are presented in Fig.\ref{fig:SubjectiveOpinion}. The figure reflects participants' opinions about the usefulness of VASSL, which increases effectiveness as observed in the first five factors. However, the figure shows significantly lower scores for VASSL in most of the ease of use factors compared to manual labeling.

\section{Discussion}
In this section, we provide more discussions on our experiment findings.

\paragraph{Subjects are not experts} Participants were college students instead of the targeted expert users due to availability. We acknowledge that the results may not represent an outcome with our intended users. However, we argue that better results would be observed if the system was tested by experts familiar with tuning clustering and dimensionality reduction methods as well as domain knowledge about spambots' behaviour. Many of the useful functionalities in VASSL were not utilized by our subjects because they were not familiar with them. For example, we noticed the common pattern of under-utilizing the tuning capabilities in the system. Subjects trusted the default values of the hyper-parameters and did not test any other alternatives that could have potentially improved their labeling performance. One can argue with confidence that an expert would not always accept the default values and would likely benefit from the flexibility of VASSL in enabling users to tune these hyper-parameters. Nevertheless, we acknowledge that testing the system with college students instead of expert users is a limitation we need to address in the future to gain more insight into other needed features as well as improvements to current features.

\paragraph{Objective results discussion} The results of the conducted user study show an improvement in the performance of labeling spambots when compared to a manual labeling procedure. However, this improvement is not statistically significant for the precision criteria. We anticipated this result, because of the way our approach works. In some cases, clustering the accounts and analyzing them as groups could lead to false positives (genuine accounts labeled as spambots). For example, most of our participants reached an insight that enabled clustering many spambots using a single feature, i.e. zero number of likes for all the tweets, indicating a spambot. However, relying on this hypothesis could include genuine accounts that are less active on Twitter, or whose tweets are not liked as often. If the size of the cluster is big, the included genuine accounts may not undergo a detailed analysis, leading to incorrect labeling. To overcome this issue, we recommend first analyzing small clusters and carefully expanding them as needed.

\paragraph{Subjective results discussion} The subjective scores collected from our participants highlighted a limitation of VASSL: although most of the participants found the tool useful, they thought it was difficult to use. Given the complexity of the system, this is mostly due to the steep learning curve required to work with the system. Participants worked with the system for only 30 minutes, after all. This finding can also be extracted from the qualitative data we have collected. The most common comment that appears in participant responses when asked about the most negative aspect in the system was \enquote{It is not too simple to use the tool.} Our focus during the design of the implementation was devoted to the utility aspects, targeting experienced analysts who regularly label Twitter spambots as our audience, so we included many functionalities that could help in various situations related to expert labeling. However, this focus on utility creates many usability limitations. One of the subjects stated, \enquote{I liked the multiple features which could have helped me accomplish my tasks, but it was challenging to learn to use them properly.}

\paragraph{Qualitative results discussion} The exit questionnaire contained open-ended questions such as \enquote{What was your policy of labeling accounts as spambots?} and \enquote{What features/functionalities in VASSL helped you in implementing your policy?} We noticed from the collected answers that most of the subjects preferred to use one view as a primary view and confirm information from another view, but they rarely used more than two views. The DR view was the most utilized view and it was commonly complemented by examining selected accounts' details from the tweets view. Another common pattern was to select one feature in the feature explorer view, and select and analyze accounts according to pre-developed hypotheses such as \enquote{Low \textit{like} counts.} We can see from these patterns the variety of supported analyses in VASSL, which provides tools that supported different workflows.

\section{Conclusion}
In this paper, we presented VASSL, a visual analytics toolkit that supports the analysis and labeling of social media accounts, for the specific use of identifying spambots. The datasets created using VASSL can be used in improved automated approaches for detecting spambots, whose nature and behavior changes dynamically to escape the current algorithms. With an effective tool such as VASSL, it is possible to quickly generate large annotated datasets that reflect the behavior of social spambots. We presented a detailed use case of the toolkit to perform a complete analysis and labeling task. Finally, we evaluated VASSL and demonstrated the significant improvement in labeling performance.

Our immediate next step is to investigate  VASSL with more machine learning models to provide suggestions for users. Similarly, user annotations should progressively improve automatic model suggestions. Future research should focus on testing the benefit of adding active learning and machine suggestion to spambot labeling. Furthermore, collaboration modules need to be created that will enable multiple users to collaborate to generate labels with minimal effort, with the help of machine-learning models as well as the insight generated by other collaborators.

\acknowledgments{
The authors wish to thank Christina Stober for proofreading the paper.}

\bibliographystyle{abbrv-doi}

\bibliography{Draft2}
\end{document}